\documentclass[%
 reprint,
 amsmath,amssymb,
 aps,
nofootinbib,twocolumn
]{openjournal}

\usepackage{aas_macros}
\usepackage{graphicx}
\usepackage{dcolumn}
\usepackage{bm}
\usepackage[mathlines]{lineno}
\usepackage{soul}
\usepackage[dvipsnames]{xcolor}
\usepackage{color}

\begin{document}


\title{$\Lambda$CDM is  alive and well
}

 \author{Alain Blanchard}
\affiliation{Universit{\'e} de Toulouse, UPS-OMP, IRAP, CNRS, 14 Avenue Edouard Belin, F-31400 Toulouse, France}

\author{Jean-Yves H\'eloret}
\affiliation{Universit{\'e} de Toulouse, UPS-OMP, IRAP, CNRS, 14 Avenue Edouard Belin, F-31400 Toulouse, France}

 \author{St\'ephane Ili\'c }

\affiliation{    Université Paris-Saclay, CNRS/IN2P3, IJCLab, 91405 Orsay, France }
\affiliation{Centre National d'\'Etudes Spatiales, Toulouse, France
}
\affiliation{
    IRAP, Universit\'e de Toulouse, CNRS, CNES, UPS, Toulouse, France}
    
\author{Brahim Lamine}
\affiliation{Universit{\'e} de Toulouse, UPS-OMP, IRAP, CNRS, 14 Avenue Edouard Belin, F-31400 Toulouse, France}

\author{Isaac Tutusaus}
\affiliation{Universit\'e de Gen\`eve, D\'epartement de Physique Th\'eorique and Centre for Astroparticle Physics, 24 quai Ernest-Ansermet, CH-1211 Gen\`eve 4, Switzerland}
\affiliation{Institute of Space Sciences (ICE, CSIC), Campus UAB, Carrer de Can Magrans, s/n, 08193 Barcelona, Spain}
\affiliation{Institut d`Estudis Espacials de Catalunya (IEEC), Carrer Gran Capit\`a 2-4, 08193 Barcelona, Spain}
\affiliation{Universit{\'e} de Toulouse, UPS-OMP, IRAP, CNRS, 14 Avenue Edouard Belin, F-31400 Toulouse, France}

\date{\today}

\begin{abstract}
Despite its successes, the $\Lambda$CDM model 
faces  several tensions with recent cosmological data and their increased accuracy. The mismatch between the values of the Hubble constant $H_0$ obtained from {some} direct distance ladder measurements and from the cosmic microwave background (CMB) is the most statistically significant, but the amplitude of the matter fluctuations is also regarded as a serious concern, leading to the investigation of a plethora of alternative models.

Here, we examine the situation from a different perspective. We first show that the combination of several recent measurements from local probes 
leads to a tight constraint on the present-day matter density $\Omega_m$ as well as on the amplitude of the matter fluctuations, both acceptably consistent with the values inferred from the CMB. Secondly, we address the Hubble tension by assuming that some determinations of the  value of $H_0$ are possibly biased. We treat such a bias as a nuisance parameter within $\Lambda$CDM 
and we examine such  ``$\Lambda$CDM$+$ $H_0$ bias'' models on the same statistical grounds as alternative cosmological  models. A bias in Planck or in SH0ES produces similar improvements. However we show that a bias in the Cepheids calibration
 produces improvements in terms of $\Delta AIC$ that supersede  
existing extended models proposed up to now. In a third step,  we show that the value of $\Omega_m$ we obtained from our RSD, Pantheon+ and 3$\times$2pt from DES, 
combined with SH0ES determination of $H_0$,
leads to a {precise determination} of the
density parameter of the Universe 
$\omega_m =  0.1753 \pm 0.0069$.
This measurement provides an additional low-redshift test for cosmological models by comparing it to the
preferred value derived from the CMB.
 From this test, most $\Lambda$CDM  extensions seem to be confronted with a new tension as many of them cluster around  $\omega_m =  0.14$, 
while there is no tension with a local   $H_0 \sim 67 $ km/s/Mpc.
We conclude that a standard $\Lambda$CDM model with an unknown bias in the Cepheids distance calibration represents a model that reaches a remarkable agreement, statistically better than previously proposed extensions without bias for which such a comparison can be performed. 
\end{abstract}

\maketitle

\section{Introduction}

\noindent The concordance model in cosmology, $\Lambda$CDM, is extremely successful in accounting for most of the current cosmological observations with great precision~\citep{2013PhR...530...87W}. 
Indeed, the accuracy of data has remarkably improved: cosmic microwave background (CMB) measurements with Planck have achieved sub-percent accuracy, reaching a precision of the order of one percent for $\Lambda$CDM parameters, providing accurate reference values for the parameters of the standard model \citep{2020A&A...641A...6P}.  After the discovery of the baryon acoustic oscillations (BAO) peak \citep{2005MNRAS.362..505C,2005ApJ...633..560E}, the properties of the 3D matter distribution in the late Universe (roughly at redshifts $z < 3$) have been sampled thanks to larger and larger galaxy (and quasars) surveys. One of the most recent instances has been provided by the final eBOSS survey, the culmination of more than twenty years of efforts \citep{2021PhRvD.103h3533A}. Moreover, large-area deep photometric surveys are now providing additional direct information on the total matter distribution through weak lensing measurements, such as the latest DES Year~3 (DES Y3) results \citep{2022PhRvD.105b3520A}. Finally, the Hubble diagram of type Ia supernovae, historically the first evidence for an accelerated expansion \citep{1998AJ....116.1009R,1999ApJ...517..565P}, is reaching high statistical power \citep{2014A&A...568A..22B,2022ApJ...938..110B}.

 On the other hand, this model also faces important problems, one of which is its inability to elucidate the mystery of the cosmological constant $\Lambda$, which remains an \textit{ad hoc} addition to general relativity, introduced to explain the late-stage accelerated expansion of the Universe. While this constant behaves in the same way as a vacuum energy, its value is many orders of magnitude smaller than the estimations of quantum field theory \citep{1989RvMP...61....1W}. Finding a solution with $\Lambda=0$ would solve one aspect of this problem, since it would be compatible with a renormalization to zero -- a more natural value compared to the current estimations of $\Lambda$. Assuming the existence of a mechanism that cancels out the contribution of the quantum vacuum, the origin of the accelerated expansion remains to be explained. This has motivated the search for alternatives to a genuine cosmological constant for the origin of cosmic acceleration, either in the form of additional fields contributing to the energy density budget, or modifications of gravity at large scales.

 Seeking viable alternatives is reassured by the emergence of what are known as 
  tensions between predictions of the standard $\Lambda$CDM (normalized from Planck) and several observational quantities. Such tensions already exist -- and were noticed -- from the CMB data as it prefers models with $w \sim -1.5 $ and a non zero-curvature $ \Omega_k \sim -0.05$ \citep{2020A&A...641A...6P}\footnote{In such cases, comparing frequentist and Bayesian approaches is useful to appreciate cosmological inferences \citep{2023PhRvD.108l3514H}.}, {although these indications are rather fragile and the most recent analyses do not support them \citep{2022MNRAS.517.4620R,2024A&A...682A..37T}}. These tensions have been regarded as a reliable indication for the necessity of alternative models \citep{2021ApJ...908L...9D,2022JHEAp..34...49A} (a broader discussion on the incompleteness of the $\Lambda$CDM can be found in \cite{2022arXiv220805018P}). This class of models is dubbed ``extensions" of $\Lambda$ models. However, combining measurements of the BAO scale and CMB data restricts considerably the kind of alternative to the standard model \citep{2006A&A...449..925B} and relatively contrived  models are needed.
 
 Additionally, during recent years tensions between several observationally-inferred cosmological parameters have emerged mostly between values inferred from early-physics data (CMB anisotropies, BAO scale) and measurements performed from the local Universe, in particular the distance ladder. From Planck data the present day Hubble constant (within standard $\Lambda$CDM) is expected to be $H_0 = 67.36\pm 0.54$ km/s/Mpc \citep{2020A&A...641A...6P}. The BAO scales measured from different redshift surveys combined with Big Bang nucleosynthesis (BBN) constraints on $\Omega_b$ allows us to tightly constrain the Hubble constant \citep{2015PhRvD..92l3516A}. Using this method the eBOSS surveys covering the redshift range $0.1 \leq z \leq 2.$ lead to $H_0 = 67.35\pm 0.97$ km/s/Mpc \citep{2021PhRvD.103h3533A}. This value conflicts with several measurements of the local expansion rate, the most discrepant being the recent values obtained by the SH0ES team \citep{2022ApJ...934L...7R} $H_0 = 73.3\pm 1.04$ km/s/Mpc. Although many disagreements have been noticed between the standard model and astronomical observations \citep{2016PDU....12...56B}, the amplitude of matter fluctuations at low redshift is regarded as the next most serious concern \citep{2021APh...13102604D}, even though the solution to it might be found in the standard model itself \citep{2022MNRAS.516.5355A}.

 In this paper, we present a different  view on the treatment of tensions. {We aim to explore the potential presence of systematics in} 
 some data, and in such cases quantify the corresponding statistical gain for the $\Lambda$CDM model in comparison to physical extensions. We first examine the so-called $S_8$ tension and then examine the possible existence of a bias affecting distance ladder measurements. Finally, we point out that the reduced matter density parameter of the Universe $\omega_m$ provides a well-measured quantity at low redshift allowing for a new test to distinguish between competitive models. 

\section{The amplitude of matter fluctuations}\label{sec2}

\noindent The amplitude of matter fluctuations is usually expressed by $\sigma_8$, i.e. in term of the average r.m.s fluctuation in a sphere of $8h^{-1}$Mpc. However, probes of matter fluctuations such as weak lensing or the abundance of clusters are sensitive to a combination of $\Omega_m$ and $\sigma_8$. This has led to the introduction of another parameter:
\begin{equation}
S_8 = \sigma_8\left(\frac{\Omega_m}{0.3}\right)^{1/2}\,,
\end{equation}
which for clusters is much less sensitive to $\Omega_m$. {However,} $S_8$ may {still} vary with $\Omega_m$ for other probes. Therefore the dependence on $\Omega_m$ is to be taken into account for model comparison.\\

\subsection{Constraining $S_8$ and ${\Omega_m}$} 

Redshift space distortions (RSD) are used to extract relevant information on the amplitude of matter fluctuations from redshift surveys{, as well as peculiar velocity field reconstruction}. The velocity field {is primarily sensitive to $
 \beta \equiv f/b$ where $b$ is the bias parameter and $f$ is the linear growth rate ($=d\ln D/d\ln a$), $D$ being the linear growth function ($ \delta(a) = \delta_0D(a)$). Results are most commonly expressed in term of the  parameterised growth rate: $f\sigma_8 \equiv f(z)\sigma_8(z)$ related to $\beta$ by the amplitude of the fluctuations in the tracer field $\widehat{\sigma}_8$: $f\sigma_8 = \beta\widehat{\sigma}_8$}. Survey analysis can therefore be used to put constraints in the  $f\sigma_8$~--~$\Omega_m$ plane \citep{2009MNRAS.393..297P,2021PDU....3100766B}. RSD measurements could be subjected to some bias in their measurement: \cite{2012MNRAS.423.3430B} showed that a cut on small scale is needed to overcome non-linear dynamic effect. Consequently RSD measurements mainly come from large scales (typically $\geq 15h^{-1}$Mpc) in the linear regime. Intrinsic alignments are also a potential problem \citep{2009MNRAS.399.1074H},  a concern for the next generation of surveys like DESI \citep{2023MNRAS.522..117L}. However, existing measurements are not always independent.
{Certain ones,} such as the eBOSS  measurements were extensively tested against mock samples. Recently, it has been shown that the eBOSS measurements, assuming Planck CMB  priors for all cosmological parameters {except}  $\sigma_8$, leads to a high value of {$S_8 = 0.841\pm 0.038$} \citep{2021A&A...656A..75B}, {consistent with recent values obtained from full shape analyses \citep{2022arXiv220608327D,2023JCAP...06..005M}. The uncertainties are large in the eBOSS RSD data, preventing the distinction between Planck and the low weak lensing estimates of $S_8$, 
as outlined by \cite{2022MNRAS.516.5355A}}. In the following, in order to achieve accurate measurements without using Planck priors, 
we use a large set of RSD measurements which we selected to be independent, following closely the approach in \cite{2022EPJC...82..594A}. To achieve this, we regard data from different surveys as being independent from each other when they cover different sky areas for the same tracer, disjointed redshift bins for the same tracer, {or}  overlapping redshift bins but with different tracers. We use the latest measurements when the same survey was analysed several times. We also remove measurements that used a fiducial model for $\sigma_8$.

\begin{table}[t]
\begin {center}
\begin{tabular}{c c c c}
    {Survey} & {$z$} & {$f \sigma_{8}$} &  {Refs}  \\
    \hline
    \hline
    2MFT & 0 & $0.505^{+0.089}_{-0.079}$&  \cite{2017MNRAS.471.3135H} \\ 
   
    6dFGS & 0.067 & 0.423$\pm$0.055 & \cite{2012MNRAS.423.3430B}\\ 
    
    SDSS DR13 & 0.1 &0.48$\pm$0.16 & \cite{2017MNRAS.468.1420F}\\ 
    
    2dFGRS& 0.17 & 0.51$\pm$0.06 & \cite{2009JCAP...10..004S}\\ 
    
    GAMA & 0.2 & 0.43 $\pm$ 0.05 & \cite{2021MNRAS.503...59A}\\ 
    
    WiggleZ & 0.22 & 0.42$\pm$0.07 &  \cite{2011MNRAS.415.2876B} \\ 
    
    BOSS LOW Z & 0.25 & 0.471$\pm$0.024 & \cite{2022MNRAS.509.1779L}\\ 




    GAMA & 0.38 & 0.44$\pm$ 0.06 & \cite{2013MNRAS.436.3089B} \\ 
    
    BOSS LOW Z & 0.4 & 0.431$\pm$0.025 & \cite{2022MNRAS.509.1779L}\\ 

    WiggleZ & 0.41 &	0.45$\pm$0.04	& \cite{2011MNRAS.415.2876B}\\ 
    
    CMASS BOSS &	0.57 &  0.453$\pm$0.022$^\dag$ & \cite{2019PhRvD.100b3504N} \\ 
    
    
    WiggleZ & 0.6 & 0.43$\pm$0.04& \cite{2011MNRAS.415.2876B}\\ 
    
    VIPERS & 0.6 & 0.48$\pm$0.12$^{\ddagger}$ & \cite{2017AA...608A..44D}\\ 
    
    SDSS IV & 0.69 & 0.447$\pm$0.039 & \cite{2020MNRAS.499.4140N}\\ 
    
       
     SDSS IV & 0.77 & 0.432$\pm$0.038 & \cite{2020MNRAS.498.3470W}\\

    WiggleZ & 0.78 & 0.38$\pm$0.04 & \cite{2011MNRAS.415.2876B}\\ 
    
    SDSS IV & 0.85 & 0.52$\pm$0.10&  \cite{2022MNRAS.513..186A}\\ 
    
    VIPERS & 0.86 & 0.48$\pm$0.10$^{\ddagger}$ & \cite{2017AA...608A..44D}\\ 
    
   SDSS IV & 0.978 &0.379$\pm$0.176$^\dag$& \cite{2019MNRAS.482.3497Z}\\
     
    SDSS IV & 1.23 &0.385$\pm$0.1$^\dag$ & \cite{2019MNRAS.482.3497Z} \\
    
    Fastsound & 1.4 & $0.494^{+0.126}_{-0.120}$  & \cite{2016PASJ...68...38O} \\ 
    
    SDSS IV & 1.52 & 0.426$\pm$0.077  & \cite{2018MNRAS.477.1639Z} \\ 
    
    SDSS IV & 1.944 & 0.364$\pm$0.106$^\dag$ & \cite{2019MNRAS.482.3497Z} \\ 
    \hline
    \hline
\end{tabular}
\caption{Selected values of $f\sigma_{8}(z)$ measured from redshift space distortion used in this work.  $^\dag$ : obtained from RSD combined with Alcock-Paczynski. $^{\ddagger}$ : for VIPERS we used their values of $f$ and $\sigma_8(z)$.}
\label{mesures}
\end{center}
\end{table}

\begin{figure}[tbp]
    \centering
    \includegraphics[width=\linewidth]{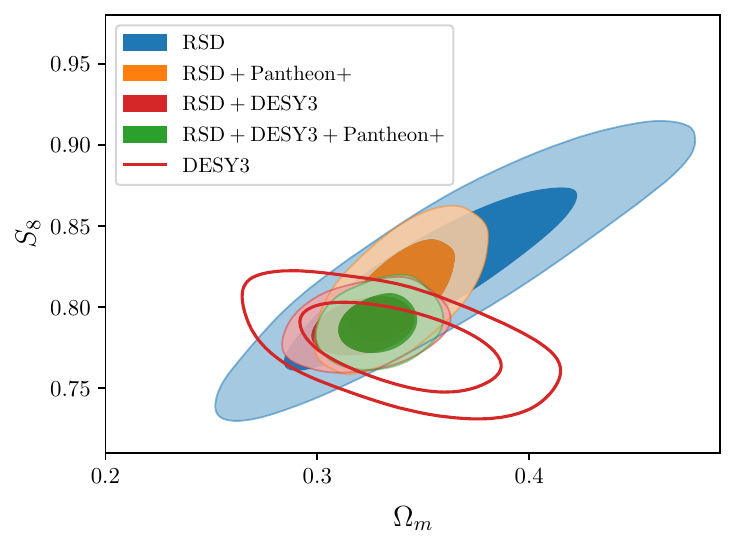}
    \caption{Confidence contours (68 and 95\%) from RSD and other data combinations. The red lines are for DES Y3 3$\times$2pt constraints. The tight final contours (green) are obtained from low-redshift data only. \\}
\label{fig:S8Om}
\end{figure}

 Our final sample  is presented in Table \ref{mesures}. 
 {Searching for the best-fit $\Lambda$CDM model, we obtained a minimum $\chi^2$ value of 11.15 (for 23 d.o.f.), $\Omega_m = 0.347$, and $S_8 = 0.816$. We also derive from this dataset the following constraints via a Monte Carlo Markov chain (MCMC): $\Omega_m = 0.349_{-0.054}^{+0.043}$ and $S_8 = {0.818}_{-0.036}^{+0.039}$.} These values and their uncertainties are similar to those obtained by \cite{2022EPJC...82..594A}. There is a clear degeneracy in the $\Omega_m$--$S_8$ plane {(blue contours in Figure~\ref{fig:S8Om})}. It is therefore convenient to combine them
 with other low-redshift probes in order to restrict the range of $\Omega_m$. We {will} use first the latest Pantheon+ result from which was derived for the $\Lambda$CDM model $\Omega_m= 0.334\pm 0.018$ \citep{2022ApJ...938..110B}, and {then} the DES Y3 3$\times$2 pt analysis \citep{2022PhRvD.105b3520A}. {Let us focus first on the} Pantheon+ measurement{, which} is interesting as it provides a tight measurement of the density parameter $\Omega_m$ from low-redshift data (in $\Lambda$CDM). When {the} {Pantheon+ {$\Omega_m$ measurement} is} combined with our RSD sample, we obtain {via MCMC} :
 \begin{eqnarray}
 \Omega_m = {0.336}_{-0.017}^{+0.017} \\
 S_8 = {0.811}_{-0.021}^{+0.021}.
 \end{eqnarray}
 Unsurprisingly, the central value of $\Omega_m$ is close to the Pantheon+ value, but the implication for $S_8$ is interesting: its uncertainties are noticeably reduced{, as expected, but also}  the central value is  {higher than current determination obtained through other approaches, as summarized in Fig. 4 by \cite{2022JHEAp..34...49A}}, making it in good agreement with Planck ($S_8 = 0.834\pm 0.016$, $\Omega_m= 0.3166\pm 0.0084$ {from} \cite{2020A&A...641A...6P}, in line with \cite{2021A&A...656A..75B}, but now obtained without priors from the early Universe {and with uncertainties comparable to  those obtained from  the best weak lensing surveys. The tight constraints obtained from the SNIa Hubble diagram by Pantheon+ and the shift in the central value obtained with Pantheon ($\Omega_m=0.298\pm 0.022$ \cite{2018ApJ...859..101S}) represent appreciable changes  for cosmological issues  (this will be illustrated  in section \ref{sec4}). This is conforted by the recent SN5YR DES results \citep{2024arXiv240102929D} obtaining very similar values ($\Omega_m=0.352\pm 0.017$)}. It should be borne in mind, however, that RSD data  are still suspected of being subject to unidentified systematics \citep{2022arXiv221116794Y}. {Our final constraint on $\Omega_m$ heavily relies on the constraints from Pantheon+ which uncertainties may be underestimated \citep{2023arXiv231112098R}. However, the  DES SN5YR combined with BAO and 3x2pt leads to $\Omega_m=0.330^{+0.011}_{-0.010}$ \citep{2024arXiv240102929D}, almost identical to our constraint below, Eq. (\ref{omega}).}

 Weak lensing surveys  are a major source of direct information on the dark matter distribution. As it is commonly the case for cosmological probes, they provide tight constraints on a combination of $\sigma_8$ and $\Omega_m$. They may however be affected by subtle unidentified biases \citep{2022A&A...664A..77S,2018MNRAS.476..151E}. The 3$\times$2 pt analysis of DES Y3 data provides a measurement of $S_8 = 0.776\pm 0.017$ \citep{2022PhRvD.105b3520A}, which is claimed to be consistent with Planck. {We perform some importance sampling of the official DES Y3 public MCMC results\footnote{https://des.ncsa.illinois.edu/releases/y3a2} by} combining  {them with our RSD dataset}, and we obtain $\Omega_m= 0.322\pm 0.016$ and a slightly higher amplitude with smaller uncertainties: $S_8 = 0.788\pm 0.012$. The three datasets (RSD, Pantheon+, DES Y3) can be assumed to be independent, as they represent different probes over different areas of the sky.
This {final combination (obtained as well through importance sampling of the DES chains)} leads to a value of $\Omega_m$ which is tightly constrained: 
\begin{equation}
    \Omega_m= 0.330\pm 0.012 \,.
    \label{omega}
\end{equation} 
We also obtained a value of
\begin{equation}
    S_8 = 0.790\pm 0.012\,. 
\end{equation}
Those results are summarized in Fig.~\ref{fig:S8Om} {and in the Fig.~\ref{fig:whisker_Omega_m} and Fig.~\ref{fig:whisker_S8} where the different measurements of $\Omega_m$ and $S_8$ are summarized}. The final  constraint on $S_8$ is in reasonable agreement with Planck CMB, 
although $S_8$ differs at a 2.2 $\sigma$ level, a conclusion similar to \cite{2021MNRAS.505.5427N} based on slightly different RSD data combined with Pantheon and BAO. Such a level of disagreement is not to be regarded as problematic in the context of model comparisons, especially given the possible non-Gaussian distribution of the measured quantities derived from weak lensing \citep{2021arXiv210710291S}. 
{The Kilo Degree Survey (KiDS) also found low value of $S_8$ with $S_8= 0.754^{ +0.027}_{-0.029}$ \citep{2022A&A...665A..56L}. There are different complex analysis pipelines of weak lensing surveys and the progress made in terms of study area has not yet been translated into similar improvements in constraining power as noticed by  \cite{2023OJAp....6E..36D}. In their joined analysis of KiDS and DES Y3, they found $S_8= 0.790^{+0.018}_{-0.014}$ with a maximum of the posterior at 0.801 and concluded to be in agreement with Planck at 1.7 $\sigma$ (see Fig.~\ref{fig:whisker_S8}).}
The various datasets and combinations used in the present work are all based on low-redshift probes. We conclude from this study that no stringent tension exists on the amplitude of matter fluctuations when comparing estimates from local probes with ones deduced from Planck in the $\Lambda$CDM model. 

\subsection{The role of baryonic  physics}


{Significant feedback is needed to prevent excessive cooling of the baryonic component of the universe, i.e.} the overcooling problem \citep{1991ApJ...367...45C,1992A&A...264..365B}. Hydro simulations have shown that the 
matter power spectrum on scales relevant for cosmological weak lensing studies is modified by the necessity of incorporating strong feedback to address the overcooling problem \citep{2011MNRAS.415.3649V,2011MNRAS.417.2020S,2019JCAP...03..020S,2023MNRAS.518..477A}. Consequently, employing power spectra from dark matter simulations can introduce considerable biases in the inferred cosmological parameters.

\begin{figure}[t]
    \centering
    \includegraphics[width=\linewidth]{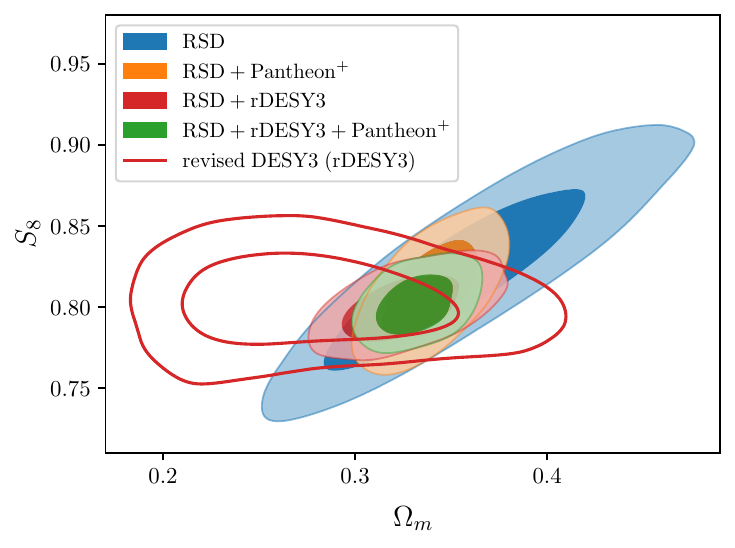}
    \caption{As in figure  \ref{fig:S8Om},  the red lines are now for the revised DES likelihood according to \cite{2023A&A...678A.109A}. \\}
\label{fig:S8Om2}
\end{figure}
{Recently \cite{2022MNRAS.516.5355A} proposed that the matter power spectrum can be suppressed on non-linear scales more than usually assumed and that this results in lowering the inferred $S_8$ from weak lensing surveys. Indeed, by modeling the baryonic physics on non-linear scales \cite{2023A&A...678A.109A} found a higher amplitude from the DES Y3 data. In order to illustrate the difference we provide an equivalent of Fig. \ref{fig:S8Om} using the revised DES Y3 likelihood. Not surprisingly, the final $S_8$ is higher,
\begin{equation}
    S_8 = 0.8009^{+0.012}_{-0.012},
\end{equation}
when combined with Pantheon+, our RSD and the revised DES Y3 likelihood revised by \cite{2023A&A...678A.109A}. The difference with Planck being now below $2\sigma$ (see Fig.~\ref{fig:whisker_S8}). The $\Omega_m$ value turns out to be
\begin{equation}
    \Omega_m = 0.3326^{+0.012}_{-0.014}\,.
\end{equation}}

\subsection{Further recent indications for a lack of tension on $S_8$}

{We notice that new recent measurements from the CMB are in line with the above conclusion on an amplitude of $S_8$ consistent with Planck CMB. In 
\cite{2023arXiv230410219R}, an estimation of weak lensing cluster mass from KiDS for a sample of SZ  clusters detected by Atacama Cosmology Telescope  (ACT) is provided, leading  to $1-b = 0.65\pm 0.05$ which translates to $S_8 \sim 0.82 \pm 0.02$. 
It is also interesting to notice that \cite{2024MNRAS.527.1244S} concluded to the consistency of field-level inference of cluster masses with LCDM, Planck normalized.}\\

{The CMB fluctuations are distinctly distorted by the gravitational  weak lensing effect due to the dark matter clustering at low redshift \citep{1987A&A...184....1B}. This CMB lensing signal  comes from structures at redshift between 0 and 5 with the kernel peaking at redshift around 2. ACT-DR6 has recently obtained a tight constrain on $S_8$ from their CMB lensing signal which has been combined with CMB lensing measured by the Planck satellite, yealding $S_8 \sim 0.831 \pm 0.023$ \citep{2024ApJ...962..113M,2024ApJ...962..112Q}.
Furthermore, using the unWISE sample of galaxies they could perform a tomographic estimate of $S_8$ with the CMB lensing signal and concluded to $S_8 = 0.813\pm 0.015$ in the redshift range  $z\approx $ 0.2--1.6 
\citep{2023arXiv230905659F}, improving the previous measurement from \cite{2021JCAP...12..028K}.
It is interesting to note that  SPT-3G obtained similar value but with larger uncertainty $S_8 \sim 0.836 \pm 0.039$ \citep{2023PhRvD.108l2005P}. We will end this census by quoting the latest results from  clusters counts:  using a non-standard scaling law evolution and  revised set of mass calibration  lead \cite{2024arXiv240204006A} to conclude that $S_8 = 0.81 \pm 0.02$; 
fitting the abundance of eROSITA  X-ray clusters abundances and using weak lensing calibrated mass  of clusters \citep{2024arXiv240208458G} leads to  $S_8 = 0.86 \pm 0.01$.\\}

{For the sake of completeness we combine the local above constraints (RSD + Pantheon+ + the mean value of DES Y3 and the revised DES Y3 + KiDS-ACT clusters + ACT + Planck lensing + unWISE):  
\begin{equation}
    S_8 = 0.8053\pm 0.0085\,. 
\label{eq:S8final}
\end{equation}
This value differs from the latest  Planck determination from \cite{2024A&A...682A..37T}, $S_8 = 0.819\pm 0.014$, by 0.8 $\sigma$ level (see last line of Fig.~\ref{fig:whisker_S8}).}

\begin{figure*}[th]
    \centering
    \includegraphics{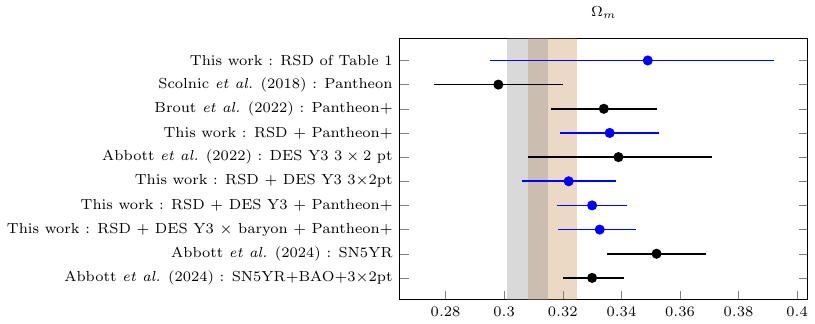}
    \caption{{Whisker plot for recent $\Omega_m$ determinations discussed in this paper. The brown band corresponds to the Planck 2018 value \citep{2020A&A...641A...6P} while the gray band corresponds to the latest PR4 Planck analysis \citep{2024A&A...682A..37T}. The DES Y3 $\times$ baryon refers to the DES re-analysis from~\cite{2023A&A...678A.109A}.}}
\label{fig:whisker_Omega_m}
\end{figure*}

\begin{figure*}[th]
    \centering
    \includegraphics{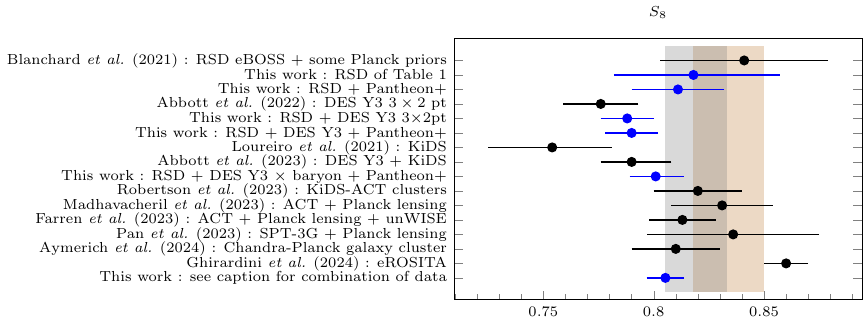}
    \caption{{Whisker plot for $S_8$ values discussed in this paper. The brown band corresponds to the Planck 2018 value \citep{2020A&A...641A...6P} while the gray band corresponds to the latest PR4 Planck analysis \citep{2024A&A...682A..37T}. The DES Y3 $\times$ baryon refers to the DES re-analysis 
    from~\cite{2023A&A...678A.109A}. The last line corresponds to the result Eq.~(\ref{eq:S8final}) corresponding to the combination RSD + Pantheon+ + the mean value of DES Y3 and the revised DES Y3 + KiDS-ACT clusters + ACT + Planck lensing + unWISE.}}
\label{fig:whisker_S8}
\end{figure*}

{The different measurements of $\Omega_m$ and $S_8$ are summarized  in the Fig.~\ref{fig:whisker_Omega_m} and Fig.~\ref{fig:whisker_S8}.}

\section{The Hubble constant}\label{sec3}

\subsection{The tension}

\noindent The tension on the Hubble constant $ H_0$ has been extensively addressed in the recent literature. {This subject is part of the long debate on the scale of distances of galaxies \citep{2023Univ....9..501C}}. The SH0ES team recently provided a value that, when compared to the Planck data, leads to the most stringent disagreement between two single measurements{, a $\sim  5\sigma$ disagreement}.
The situation worsens because on one hand several measurements of the Hubble constant from local probes lead to values consistent with the SH0ES value, 
while on the other hand the BAO length scale in combination with primordial nucleosynthesis and the SNIa Hubble diagram leads to a value consistent with Planck's one \citep{2015PhRvD..92l3516A}. Using this {BAO} approach, the recent eBOSS analysis concluded to $H_0 = 67.35 \pm 0.97$ km/s/Mpc~\citep{2021PhRvD.103h3533A}. The inconsistency of the values can be measured from the combination of the most illustrative Hubble constant measurements: in the following we used the value derived by the  Carnegie–Chicago Hubble Project (CCHP) based on the tip of the red giant branch {(TRGB)} to calibrate the distance ladder, measuring $H_0 = 69.6 \pm 1.9$ km/s/Mpc \citep{2020ApJ...891...57F}. We also use the value obtained by the Megamaser Cosmology Project (MCP) leading to $H_0 = 73.9 \pm 3.0$ km/s/Mpc \citep{2020ApJ...891L...1P}, as well as three other recent measurements: the Tully–Fisher method (TF) leading to $H_0 = 75.1  \pm 3.0$ km/s/Mpc \citep{2020ApJ...902..145K}, the surface brightness fluctuation method (SBF) leading to $H_0 = 73.7  \pm 2.4$  km/s/Mpc \citep{2021ApJ...911...65B} and the method based on Mira variables, combined with additional calibration, leading to $H_0 = 73.3  \pm 4$ km/s/Mpc \citep{2020ApJ...889....5H}. Time-delay in gravitational lenses is also often quoted, but their inferred value is known to depend on the mass modelling \citep{2022A&A...667A.123S}, preventing it from being used efficiently for now. {The Sunyaev-Zeldovich method is a historically established technique that is independent of a standard ruler. A recent determination claims good accuracy: 
$H_0 = 67  \pm 3$ km/s/Mpc \citep{2019A&A...621A..34K} } Gravitational wave (GW) events are promising technique, but the most stringent value obtained up to now is $H_0 = 67.6  \pm 4.2$ km/s/Mpc \citep{2020arXiv200914199M}.  Quoted uncertainties are often dominated by systematic effects. It is interesting to notice that the TF and SBF  methods rely on Cepheids calibration, as does the SH0ES approach. 
Taking the measurements from Planck, BAO, SH0ES, CCHP, MCP, TF, SBF and Miras the statistical average  is $H_0 = 68.8 \pm 0.4 $ km/s/Mpc (due to the high weight of the Planck value), but with a $\chi^2 \sim 37$ {for 7 degrees of freedom,} which is high, indicating the strong inconsistency between the values. 
This is commonly interpreted as the signature of a tension between values derived from early-physics considerations\footnote{The consistency of values derived from early physics has been recently reinforced by \cite{2022PhRvD.106f3530P} who derived $H_0 = 64.6 \pm 2.4$ km/s/Mpc from a combination of late-Universe data used to measure the scale associated to the horizon at matter-radiation equality, in a sound horizon-independent way{, a conlusion disputed by \cite{2023PhRvD.108j3525S}}.} and present-day values derived from the distance ladder \citep{2019NatAs...3..891V,2022JHEAp..34...49A}, given their concordance. However the situation is not entirely clear: from Cepheids and TRGB-calibrated SNIa, one of the reported determination of $H_0$ was $63.7 \pm 2.3$km/s/Mpc \citep{2013A&A...549A.136T}. This value, as well as the one based on GW is not used in the following.  Finally, {let's consider another probe, namely} the cosmic chronometers data (CC), based on the differential age of passively evolving galaxies. {It}  can be used to determine the Hubble rate at different redshifts and thereby to estimate cosmological parameters including  the Hubble constant \citep{2002ApJ...573...37J}. This method has been used in the past to  infer $H_0$  \citep{2018JCAP...04..051G}, but the accuracy of this method is limited by the uncertainties on other cosmological parameters, including the dark energy equation of state. However, {our objective here is to assess} $H_0$ for a $\Lambda$CDM model restricted to matter density inferred from Pantheon$+$: $\Omega_m = 0.334 \pm 0.018 $ \citep{2022ApJ...938..110B}. To do so, we use the latest compilation from \cite[Table 1]{2021ApJ...908...84V} and we fit the Hubble constant:
\begin{equation}
    H(z) = H_0\left(\Omega_m (1+z)^3 + 1-\Omega_m\right)^{1/2}\,.
\end{equation}
The narrow range of values of $\Omega_m$ from Pantheon+ \citep{2022ApJ...938..110B} is added as a prior and allows us to derive an accurate value of the Hubble constant, $H_0 = 67.4  \pm 1.34 \textrm{ km/s/Mpc}$,
in good agreement  with {the} value inferred from early physics within the $\Lambda$CDM model. { \cite{2020arXiv201110559R} obtained a value with a similar precision in a more general case than the $\Lambda$CDM model.} When added to the above set of Hubble constant  determinations, this value does not change any of the previous conclusions, the values of $H_0=68.6 \pm 0.4$ km/s/Mpc and of the $\chi^2\sim 38$ are almost unchanged.

However, because ages are derived from the complex modelling of the star population synthesis and stellar population properties, one may be concerned by the existence of a possible bias. Indeed, estimation of possible astrophysical uncertainties in the modelling leads to wider error bars $H_0 = 66.5  \pm 5.44 \textrm{ km/s/Mpc}$  \citep{2022LRR....25....6M}, making this determination of little weight and is therefore not used in the following. Such  concern about possible unidentified bias should also be applied to the other Hubble measurements \citep{2021ApJ...919...16F}. Clearly, the necessity for new physics is required only if one is confident that there is no significant systematic error left over \citep{2022JHEAp..34...49A}.
Indeed the possibility of a bias in the $H_0$ determination is seriously taken in consideration by some authors in the field \citep{2021ApJ...919...16F,
2021jwst.prop.1995F}. For instance, \cite{2022ApJ...933..212M}
claims that Cepheids calibration systematics may shift down the Hubble constant by $5$ km/s/Mpc, which would essentially remove the discrepancy between the SH0ES value and the early physics ones. Recently, it was proposed that a different prior on dust extinction could lead to a shift in the preferred color correction for SN and thereby on the value of the Hubble constant \citep{2022MNRAS.515.2790W}. {However, i}t is {often} argued that the various methods used to derive the Hubble constant provide evidence that the tension is real and cannot be attributed to a possible bias in a single measurement. Hence, a plethora of models have been proposed in order to solve this issue, see for instance the impressive compilation from  \cite{2021CQGra..38o3001D} although almost none are actually viable solutions to solve the
 Hubble tension. 

\subsection{Methodology}

There are many tools proposed to quantify the respective merit of different models when compared to data, in particular in situations where ``tensions exist". A standard way to compare the merit of these models is to compute the amount by which their (best-fit) $\chi^2$ on the data improves compared to the $\Lambda$CDM model and/or examine the improvement provided in terms of the Akaike Information Criterium (AIC):
\begin{equation}
\Delta \textrm{AIC} = \Delta \chi^2 +2\Delta p \,.
\label{AIC}
\end{equation}
{where $p$ is the number of parameters.} This metric allows us to compare the improvement in $\chi^2$ but penalizes the addition of new parameters (the $\Delta p$ term). This criteria however is not necessary sufficient to decide between models \citep{2024MNRAS.tmpL..29C}.
In the present work, {our aim is to undertake a statistical comparison considering a potential systematic effect in one of the measurements.}  {To achieve} this, we consider the presence of an unknown parameter $\alpha$ that allows one observed measured quantity to shift. We determine this parameter by minimizing the $\chi^2$:
\begin{equation}
    \chi^2 = \sum_i \frac{(H_0-\alpha_i\times H_{0,i})^2}{\sigma_i^2}
    \label{chi2}
\end{equation}
$\alpha_i $ is set to 1 except for measurements that we regarded as possibly biased, for which $\alpha_i = \alpha $ is our single free parameter { i.e. each line in table \ref{table:ta} gives the measurement for which a bias $\alpha$ is allowed (two for the last line).} $H_{0,i}$ are the measures used and $\sigma_i$ their uncertainty. 
{table \ref{table:ta}  provides} the  $\chi^2$ and improvements in terms of AIC. Treating such  a possible bias as an additional parameter of the $\Lambda$CDM  allows to consider  it as an ``extension" of the basic model and to make a statistical comparison.

\subsection{Results}

In our models, named E1 to E{4}, we respectively assume the presence of an unknown bias{, $\alpha$ in Eq. \ref{chi2}, } in i) the SH0ES data {(E1)}, ii) the Cepheids calibration thereby affecting in an identical way SH0ES, TF, and SBF data, {(E2)} iii) $H_0$ from BAO data {(E3)}, iv) $H_0$ from Planck data {(E4)}. In the last one, two unknown biases are assumed, on $H_0$ from Planck and on BAO data {(E5)}. Although there is some debate on possible bias in $H_0$ distance ladder measurements \citep{2020arXiv200710716E,2024IAUS..376....1F,2023JCAP...11..050F,2024MNRAS.529.2627M,2024arXiv240112964M},  
it is generally considered not very likely for  Planck’s inferred value for $H_0$ in $\Lambda$CDM to suffer from an observational bias. Note that the final values of Planck are in agreement with WMAP+ ACT \citep{2020JCAP...12..047A} as well as with the lensing on the CMB \citep{2022JCAP...09..039C,2024ApJ...962..113M} which is observationally very different in its potential systematics.  
The results are summarized in Table~\ref{table:ta}. From this table, one can compare improvements using a Jeffrey's scale: a value of $\Delta$AIC below -10 is considered to be very strong evidence for the proposed model over the baseline \citep{2013JCAP...08..036N}. From our results, it appears that a bias in any single determination of the Hubble constant -- except from BAO (model E3) -- produces such strong evidence. 
In other words, the determination of the Hubble constant from the BAO does not change appreciably the level of disagreement among the other determinations. On the contrary, if  one assumes that SH0ES is affected by some unknown bias (model E1),  the $\Delta$AIC  is -17.7, an evidence which is very strong 
 and performing better than any model examined in \cite{2022PhR...984....1S}. However, the value of the $\chi^2$ falls to $\sim 17.3$ which remains not particularly good for seven degrees of freedom. This supports the qualitative impression that removing SH0ES does not bring a particularly good agreement between the remaining determinations. A bias in any other single low-redshift determination of $H_0$ does not produce any appreciable improvement{, while assuming a bias in $H_0$  from Planck (model E4) produces an improvement comparable to a bias in SH0ES.} 
{Overall, }few models among the extensions already investigated in the past reach such level of improvement. Model E5 is a model in which we assume that both CMB and BAO are affected by a bias  and are therefore removed from Eq. \ref{chi2}. This means removing the early universe estimations of $H_0$, and of course there is no longer any tension. The improvement is then similar to the one obtained in model E2: $\Delta$AIC $\sim -30$. 

\begin{table}[t]
\begin{center}
    \begin{tabular}{l c c }
        \hline
        \hline
        Bias allowed in: & $\chi^2$  & $\Delta $AIC \\
        \hline
        None  &  37.0 & --~ ~ \\
        SH0ES (E1)  & 17.3 &  --17.7 \\
        Cepheids (E2)  & 6.7 &  --26.3 \\
        $H_0$ from BAO (E3) & 34.4 &  --0.6 \\
        $H_0$ from Planck (E4)  & 19.2 & --15.76 \\
        $H_0$ from Planck+BAO (E5)  & 4.3 &  --30.7 \\
        \hline
        \hline
    \end{tabular}
    \caption{Quantitative success of the $\Lambda$CDM when some (unknown) bias is assumed for one (two for E5)  $H_0$ measurements among the set used (Planck, BAO, SH0ES, CCHP, MCP, TF, SBF, Miras).  }
    \label{table:ta}
    \end{center}
\end{table}

\noindent However, as mentioned earlier several methods used to infer the Hubble constant rely on the Cepheids calibration. Therefore a bias in this calibration would affect in a similar way {\em all} these determinations. Assuming such a bias (model E2) would affect not only SH0ES but also the surface brightness method as well as the Tully-Fisher one. Under this hypothesis, still adding only one parameter to the $\Lambda$CDM, we obtain a $\chi^2$ of 6.7 and a $\Delta$AIC of -26.3. 
Recently several review papers addressed various $\Lambda$CDM extensions  that may relieve the $H_0$ tension among which  \cite{2021CQGra..38o3001D,2021APh...13102604D}. \cite{2022PhR...984....1S} performed a  quantitative investigation of the performance of various $\Lambda$CDM extensions. In this work, improvements in models were examined through three distinct metrics: the Gaussian Tension (GT), the change in the effective $\chi^2$, and the improvement in  AIC. We can therefore compare directly and quantitatively the improvement from our Table 2 to the models they investigated\footnote{Because our $\chi^2$ is derived directly from $H_0$ measurements and uncertainties, the GT is less relevant  in our case.}.
 
We conclude from this comparison that the possibility of a bias in the Cepheids calibration leads to an improvement in $\chi^2$ (-30) and in  AIC (-26.3) that no extension of the $\Lambda$CDM model investigated in \cite{2022PhR...984....1S,2023arXiv231209814R} has reached. {It should be noted that only a model that entirely avoids the constraints imposed by CMB and BAO (E5) would be capable of achieving a comparable level of improvement. This is notoriously difficult  as a simple reduction the cosmic sound horizon can not solve the problem \citep{2021CmPhy...4..123J}}. It will interesting to see in the future if any 
theoretical model can reach such a level of improvement. 


\section{More tension than $H_0$:  the {$\omega_m$} {tension}.}\label{sec4}

The physical density of the Universe is measured by the parameter $\omega_m$:
\begin{equation}
    \omega_m = \Omega_m h^2\,,
    \label{PoM}
\end{equation}
which is accurately determined from Planck data:
\begin{equation}
    \omega_m = 0.1424 \pm 0.0012\,.
    \label{Pom2}
\end{equation}
{The} accuracy of $0.85\%$ is strictly valid only for $\Lambda$CDM. The observed CMB angular power spectrum ($C_\ell$) is primarily the result of the physical evolution of fluctuations in the early Universe and the distance to the last scattering surface (LSS) \citep{1984A&A...132..359B}. The early physics depends on physical densities of the various fluids, i.e. photons, baryons, and dark matter, and on the expansion rate at that time.
With standard physics in the early Universe (before LSS) the density in radiation is well known and consequently the densities of the other fluids are accurately constrained. {Increasing the dark matter content decreases the amplitude of the peaks and slightly modulate them, while changing the baryon content modulates the peaks in a specific way, odd numbered peaks are enhanced over the even numbered peaks. The exquisite Planck data on the shape of the $C_\ell$ restricts the parameters $\omega_{DM}$ (dark matter content) and $ \omega_{b}$ (baryonic content) to  very narrow ranges and thereby $\omega_{m}=\omega_{b}+\omega_{DM}$. The distance to the CMB is then tightly constrained by the angular position of the features in the $C_\ell$.} ``Early cosmological parameters", including $\omega_m$, can therefore be constrained without assumptions on the late-time evolution: \cite{2010JCAP...08..023V} {and} \cite{2013JCAP...02..001A} found uncertainties two times larger than with the standard model, staying within the 1\,$\sigma$ interval of the latter. { A more recent analysis from \cite{2023PhRvD.107j3505L} concluded that early-LCDM is~:
\begin{equation}
  \omega_m = 0.14143 \pm 0.0013\,.
   \label{OmRplanck}
\end{equation}}
The other ingredient, the distance to the LSS, is also well constrained but provides a quantity in which the Hubble constant is degenerate to first order. This is illustrated by the behavior of $\omega_m$ versus $w$ in $w$CDM ({see for instance Fig.~2 of \cite{PhysRevD.103.063539} where} it is clear that $\omega_m$ constraints are nearly independent of $w$). This degeneracy prevents an accurate determination of $H_0$ from  the CMB only.  In extended models, it is therefore natural to expect that the matter density of the Universe $\omega_m$ remains tightly constrained given the observed {shape} $C_\ell$, {intimately linked to the sound horizon scale,} while the Hubble constant may be more sensitive to the details of the model. Extensions of $\Lambda$CDM that provide a good fit to the observed $C_\ell$ are therefore expected to have a (total) matter density close to the above value.
\begin{figure}[tbp]
    \centering
    \includegraphics[width=\linewidth]{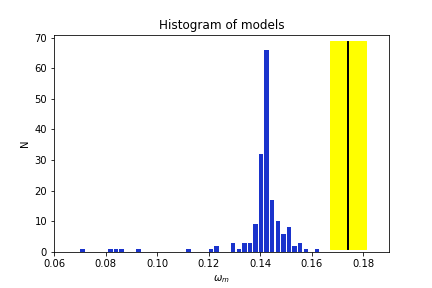}
    \caption{Histogram of the preferred value of $\omega_m$ for a series of $\Lambda$CDM extensions, adjusted to CMB data (see \citep{2021CQGra..38o3001D} for the comprehensive list). The black vertical line is our $\omega_m$ estimation for $\Lambda$CDM models from local data (\ref{omegaR}), the yellow area corresponds to the 1\,$\sigma$ region.}
    \label{histo}
\end{figure}

To check this in a more quantitative way, we use the compilation of references from \cite{2021CQGra..38o3001D} to produce the histogram of $\omega_m$ values obtained from all models when fitting the Planck $C_\ell$ (and using the FIRAS determination of the value of the temperature of the CMB\,\footnote{\cite{2020PhRvD.102f3515I} have investigated the consequences of  relaxing the temperature of the CMB from the FIRAS determination, and found a best-fit value of $H_0 \sim 55$ {km/s/Mpc} and $\Omega_m \sim 0.75$, i.e. $\omega_m \sim 0.23$. The inclusion of BAO data brings back the density to $\omega_m=0.1394$.}). This histogram is shown in Fig. \ref{histo}. 
As anticipated, values of $\omega_m$ are clustered around the best $\Lambda$CDM value. Their average and standard deviation are $0.1409 \pm 0.011$, similar to the Planck determination ({Eq.} \ref{Pom2}). This is coherent with the fact that $\omega_m$ is robustly constrained from the CMB in a relatively model-independent way. {Interestingly enough,} some points of Fig. \ref{histo} lie on the low $\omega_m$ side (corresponding to some interacting dark energy models), {while no model lie on the high $\omega_m$ side}). Furthermore, the recent sound horizon-independent constraints \citep{2022PhRvD.106f3530P} obtained from large-scale structure data in the late Universe are leading to
\begin{equation}
    \omega_m = 0.1425 \pm 0.007\,.
\end{equation}
This measurement is independent and complementary to the CMB. Another recent estimation of cosmological parameters from early-time physics (at matter-radiation equality epoch) from the full transfer function yields $\omega_m\sim 0.14$ with small uncertainties \citep{2022JCAP...08..024B}.

\noindent These values can be compared to the one we can derive from our estimate of $\Omega_m$ {(Eq. \ref{omega})}, using the SH0ES determination of the Hubble constant $H_0 = 73$ km/s/Mpc:
\begin{equation}
  \omega_m = 0.1753 \pm 0.0069\,.
   \label{omegaR}
\end{equation}
This value is more than $4\sigma$ away from Planck preferred value, Eq. \ref{Pom2}. Using $H_0 = 67$ km/s/Mpc yields $\omega_m = 0.1447 \pm 0.0058$, a value higher but fully consistent with the  Planck derived value\,\footnote{In other words the Planck central value of $\omega_m$ and the Pantheon+ determination of $\Omega_m$ point towards $H_0= 65\pm2$ km/s/Mpc in $\Lambda$CDM.}. 
The Pantheon+ 1$\sigma$ region is shown as a yellow band in Fig.~\ref{histo}. Clearly, no model sits in the preferred region of SH0ES+Pantheon+, and most fall outside of its 3$\sigma$ range. Because the late-time expansion of these models does not necessarily follow $\Lambda$CDM (i.e. as produced by a  dark energy fluid with $w = -1$), a fully quantitative comparison cannot be performed. However, the level of disagreement is so strikingly large that one may conclude it will be difficult for this type of models to be in agreement with the value of $\omega_m$ obtained with the combined SH0ES$+$Pantheon+ data. {The discriminating power of this can be seen by examining the sophisticated models investigated by \cite{2023PDU....4201348P}. Models which are fitted to cosmological data including SH0ES hardly reached 
$\omega_m = 0.1525$ which is still 3$\sigma$ lower than Eq. \ref{omegaR}. Identically the models including of magnetic fields regarded as able to solve the Hubble tension (model M1 in their table 1) leads, to $\omega_m = 0.1434\pm 0.0014$, lying more than 4$\sigma$ away from Eq. \ref{omegaR}.}\\

\section{Discussion and conclusion}\label{sec6}

\noindent In recent years, the $\Lambda$CDM model has been regarded as being in tension with the Hubble constant measurements in the local Universe, as well as with the amplitude of matter fluctuations although at a lower statistical level of significance. One convenient way to quantify the  merit of alternative models is through two  metrics{,namely} the improvement in the  $\chi^2$ and in the AIC.  With the most recent data on the amplitude of matter fluctuations from RSD data, we show that some difference between values inferred from the CMB and values inferred from local probes persists but at a moderate level of disagreement. 
The tension on the value of the Hubble constant is the most statistically significant one. {The paper} 
\cite{2022PhR...984....1S} provides a comprehensive investigation of the relative merit of extended models using  the above-mentioned metrics. It is  currently recognised that there is no satisfying physical solution to the Hubble tension \citep{2023arXiv230109695L}. It remains possible that the tension  is due to an incompleteness of the $\Lambda$CDM or to some unidentified systematics  \citep{2021A&ARv..29....9S}. 

This {situation}
clearly calls for considering the option of a possible bias in the values inferred from local data. For instance, \cite{2023MNRAS.525.5187W} claim to have identified two populations in the SNIa population, with possible consequence on the Hubble constant determination. We conducted a 
comparison by statistically addressing any potential bias as a nuisance parameter, and employing the same metrics as in \cite{2022PhR...984....1S}
(Eq. \ref{AIC} and \ref{chi2}).

{The results are} summarized in our Table \ref{table:ta}.  We found that 
models with {a} bias in $H_0$ lead{s} to a highly significant improvement in $\chi^2 $  as well as in AIC compared to existing proposed $\Lambda$CDM extensions. If one assumes that such a bias is on the Cepheids distance scale, affecting several of the Hubble constant measurements, the gain in AIC supersedes all published models. On the other hand, the remarkable agreement between distance measurements determined from HST and JWST \citep{2024ApJ...962L..17R} does not leave much room for any miscalibration.  Clearly  if the SH0ES value is essentially correct, any new physics required must be hidden in the other relevant cosmological data. 
 
Finally, introducing a new ingredient for model comparison, we showed that most models from a compilation of nearly 200 extensions lead to a value of the {physical} density parameter $\omega_m$ at odds with the value inferred from SH0ES+Pantheon+ or RSD+DESY3+SH0ES+Pantheon+  (although a direct, fully {consistent} comparison remains to be done).
{Indeed,  despite the large variety of models proposed in recent years, we have found that no model could accommodate the low redshift high value of $\omega_m$  inferred from RSD+DESY3+SH0ES+Pantheon+. }
We conclude that the $\Lambda$CDM model remains in better agreement with data than a large class of extended models that were proposed  in recent years, and that it is more likely that the local distance scale is affected by some  unidentified bias as no existing model solves the CMB and BAO scale with the same gain in AIC. The 
case for the $\Lambda$CDM model plus some bias in the Cepheid calibration is statistically preferred over the cases with extended models (from a large class proposed in recent years) plus no Cepheid bias. 

\begin{acknowledgments}
\noindent We thank G.~Efstathiou,  J.~Lesgourgues, M.~Moresco and A.~Riess for valuable  comments on the manuscript. SI thanks the Centre national d'études spatiales (CNES) which supports his postdoctoral research contract. {We are grateful to R.Angulo and G. Aric\`o for providing us with their likelihood used in Fig. \ref{fig:S8Om2}}. IT acknowledges funding from the European Research Council (ERC) under the European Union's Horizon 2020 research and innovation programme (Grant agreement No. 863929; project title ``Testing the law of gravity with novel large-scale structure observables'').
\end{acknowledgments}

\bibliographystyle{mnras}

\bibliography{Biblio.bib}

\end{document}